\input{aipcheck}

\documentclass[
    ,final            % use final for the camera ready runs
  ]
  {aipproc}

\layoutstyle{8x11single}

\usepackage{url}
\usepackage{graphicx}
\usepackage{multirow}

\begin{document}

\title{Radio Detection of Horizontal Extensive Air Showers with AERA}


\author{Olga Kambeitz$^\ast$ for the Pierre Auger
Collaboration$^\dagger$}{
  address={$^\ast$Institut f\"ur experimentelle Kernphysik, \\Karlsruhe Institute of Technology (KIT), Germany\\
        E-mail: {olga.kambeitz@kit.edu} \\
     $^\dagger$Pierre Auger Observatory, Av. San Mart\' in Norte 304,
5613 Malarg\"ue, Argentina \\ (Full author list:
\url{http://auger.org/archive/authors_2014_06.html})}
}

\begin{abstract}
AERA, the Auger Engineering Radio Array, located at the Pierre Auger Observatory in Malarg\"ue, Argentina measures the radio emission of extensive air showers in the 30-80 MHz frequency range and is optimized for the detection of air showers up to 60$^{\circ}$ zenith angle. In this contribution the motivation, the status, and first results of the analysis of horizontal air showers with AERA will be presented.
\end{abstract}

\maketitle

%%%%%%%%%%%%%%%%%%%%%%%%%%%%%%%%%%%%%%%%%%%%
%% MAINMATTER
%%%%%%%%%%%%%%%%%%%%%%%%%%%%%%%%%%%%%%%%%%%%
\section{Introduction}
AERA~\cite{AERA}, the Auger Engineering Radio Array, was designed to explore the radio emission of extensive air showers above an energy threshold of $10^{17}$ eV in a hybrid detection mode. It consists of 124 antenna stations of which 24 are log periodic dipole antennas (LPDAs) and 100 are butterfly-type antennas. Both antenna types measure two electric field components (North-South and East-West). AERA is able to measure cosmic rays together with the Auger surface detector, the fluorescence detector, as well as the muon detector of AMIGA. It has been proven that there is a correlation between the radio amplitude and the energy of the primary cosmic ray~\cite{qader_proceed}. Furthermore the correlation between radio parameters and the shower maxima has been shown~\cite{qader_proceed}.
  Taking horizontal air showers into account will increase the exposure of radio hybrid measurements as well as give better understanding of the electromagnetic component of inclined showers. This will contribute to the knowledge gained by analysing the data of the surface detector, the fluorescence detector, and the AMIGA muon detector extension. \iffalse All together it will make the Pierre Auger Observatory a powerful instrument to detect cosmic rays in all solid angles and therefore allow to detect new phenomena in cosmic rays.\fi

\section{Motivation to study horizontal showers}
\begin{figure}[b]
  \includegraphics[height=.13\textheight]{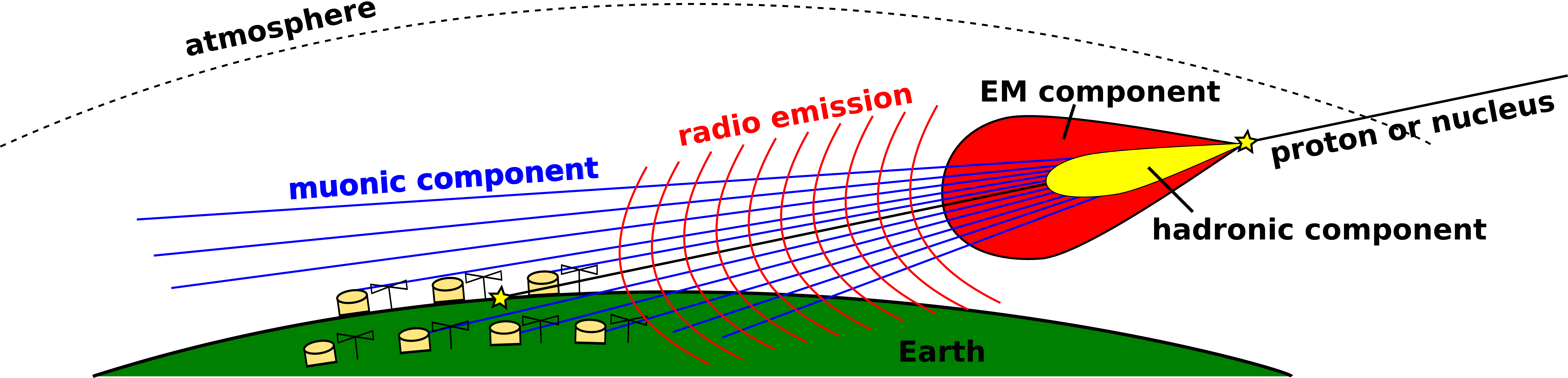}
  \caption{Illustration of a horizontal shower. The hadronic and electromagnetic component is absorbed and the muonic component and the radio signal can be measured on the ground.}
  \label{fig:sdrdinclined}
\end{figure}

\begin{figure}[!htbp]
  \includegraphics[height=.21\textheight]{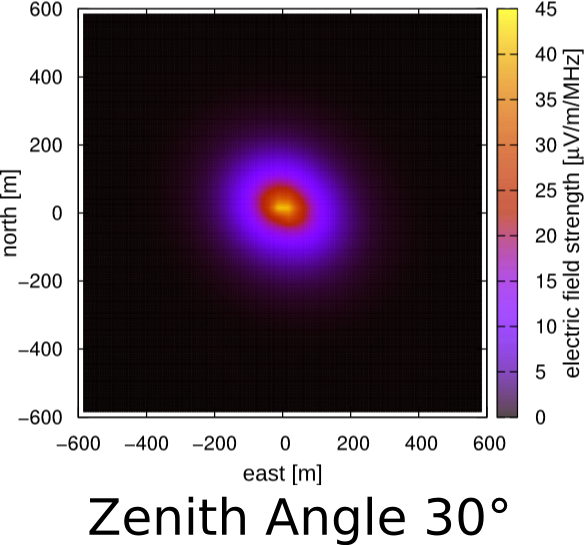}
  \includegraphics[height=.21\textheight]{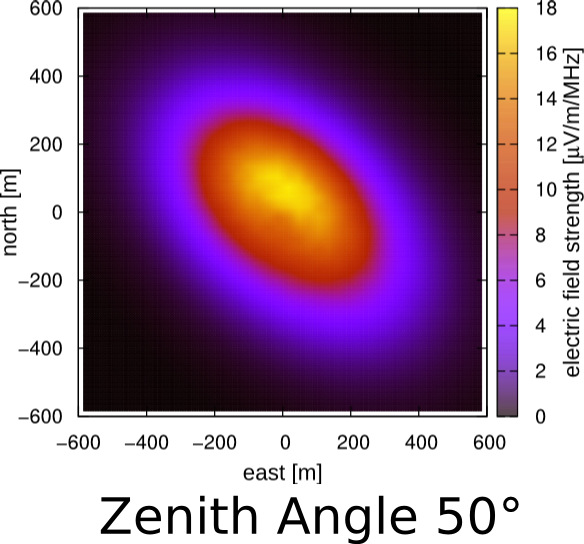}
  \includegraphics[height=.21\textheight]{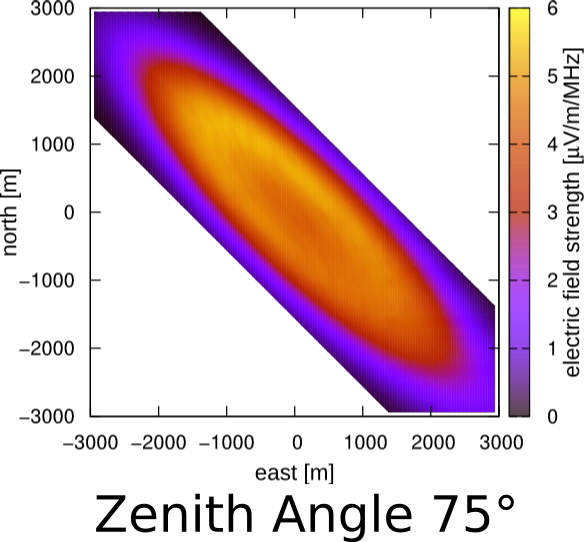}
  \caption{CoREAS simulations for primary proton showers of $10^{18}$ eV for different zenith angles~\cite{CoREAS_sim}. Please note the different scales. The limit of detection for current experiments is \(\approx\) 2 \(\mu\)V\(/\)m\(/\)MHz.}
  \label{fig:simtimewa}
\end{figure}

Showers with zenith angles larger than 60$^{\circ}$ are defined as horizontal for the surface detector infill area. \iffalse The analysis of the radio signal of horizontal showers offers some advantages.\fi The motivation to investigate horizontal air showers in the radio domain is that the radio signal is not absorbed in the Earth's atmosphere. This means, the radio signal can still be detected when the particle shower is mostly absorbed. This is illustrated in figure~\ref{fig:sdrdinclined}. 
The primary particle interacts in the atmosphere, the hadronic and electromagnetic components are evolving in the atmosphere and are then absorbed. What reaches the ground is the muonic component of the shower and the coherent radio emission emitted by the electromagnetic component.
Horizontal extensive air showers travel through more atmosphere than vertical showers, i.e. the distance to the radio source is larger. During the whole evolution of the shower, radio waves are emitted mainly in forward direction and this leads to a large footprint on ground for more horizontal showers. This is illustrated by CoREAS~\cite{CoREAS} simulations seen in figure~\ref{fig:simtimewa}. 
Three simulated extensive air showers with an energy of $10^{18}$ eV are shown with zenith angles of 30$^{\circ}$, 50$^{\circ}$ and 75$^{\circ}$. The footprint of the shower with 75$^{\circ}$ zenith angle is of 6 km length, where the asymmetry of the footprint has to be taken into account. The lateral distribution function and the wave front shape of the radio signal of horizontal air showers will provide information to distinguish between young and old showers. 
This difference is already seen just by the size of the footprint. The background and transient noise is larger for radio detection of horizontal cosmic showers because looking to the horizon many man-made noise sources are relevant. A further relevant characteristic is that a significant vertical component of the electric field vector of the radio signal is expected for the Argentinian local magnetic field.

%%%%%%%%%%%%%%%%%%%%%%%%%%%%%%%%%%%%%%%%%%%%%%%%%%%%%%%%%%%%%%%%%%%%%%%%%%%%%%%%%%%%%%%%%%%%%%%%%%%%%%%%%%%%%%%%%%

\section{Horizontal events measured by AERA-24 and AERA-124}
\begin{figure}[!htbp]
\centering
\begin{minipage}[b]{.5\linewidth}
  \begin{center}
  \includegraphics[width=.8\linewidth]{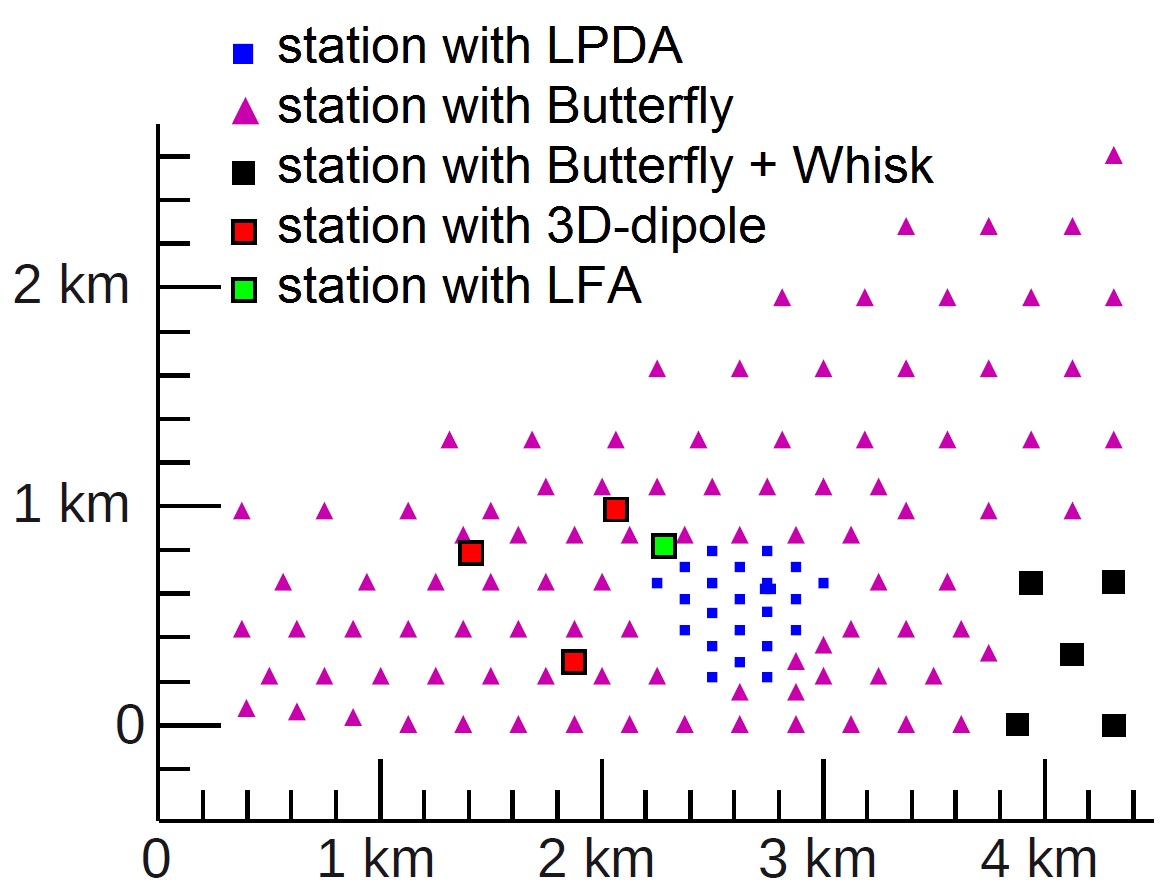}\\
  \begin{flushleft}
    \caption{figure}{Deployment of prototype stations in November 2013:
The current AERA-124 configuration with 24 butterfly stations, 100 LPDAs, five butterfly antennas with whisk antennas, three tripole stations, and one low-frequency antenna. Figure adapted from~\cite{map}.}
  \end{flushleft}
  \label{fig:aeramap}
  \end{center}
\end{minipage}%
\quad
\begin{minipage}[b]{.5\linewidth}
  \begin{center}
\includegraphics[width=.8\linewidth]{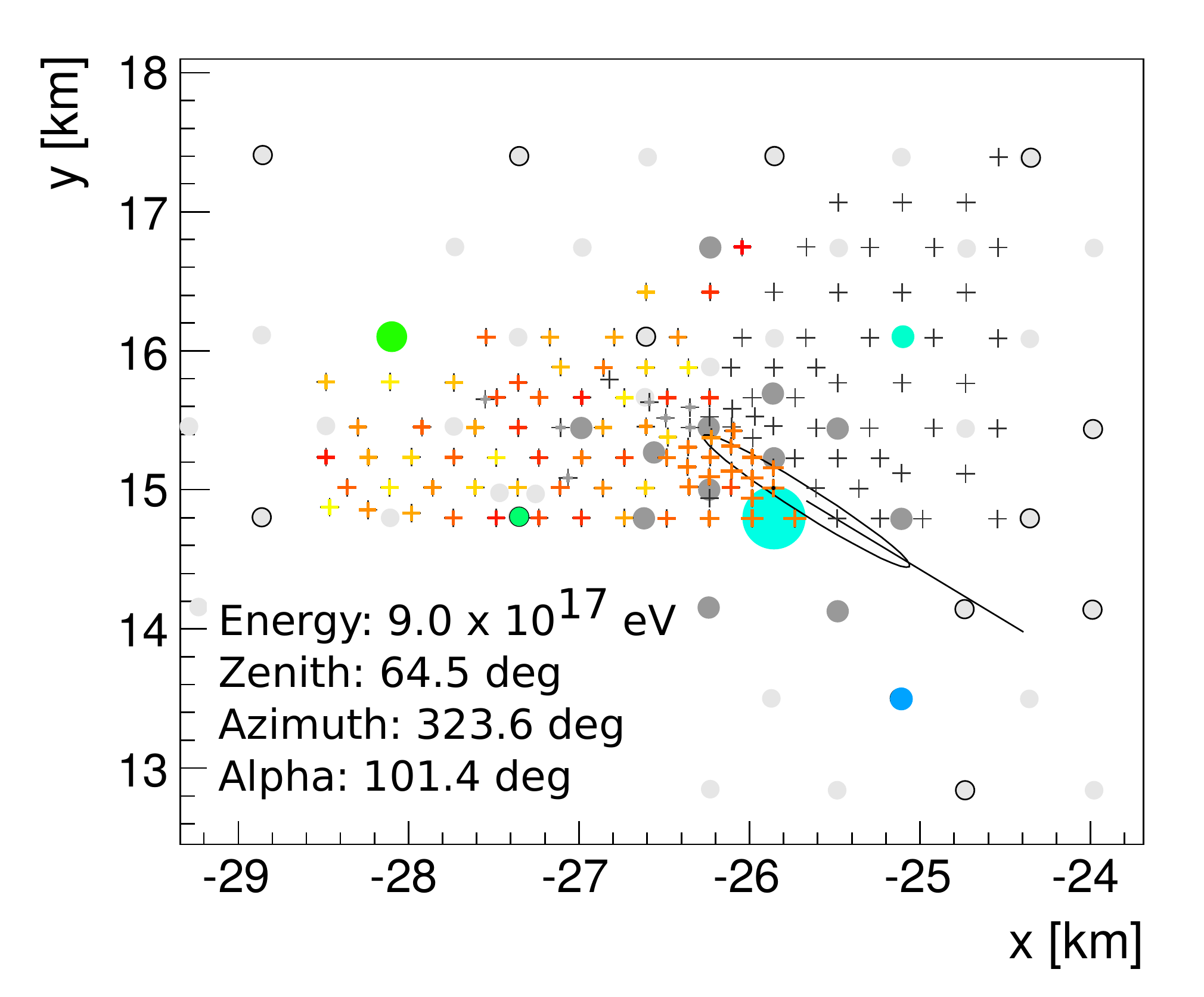}
  \begin{flushleft}
\caption{figure}{Event example of a horizontal air shower with an energy of 9.0$\times$10$^{17}$eV, zenith of 64.5$^{\circ}$, azimuth of 323.6$^{\circ}$ and geomagnetic angle alpha of 101.4$^{\circ}$.}
  \end{flushleft}
\label{fig:event_example}
  \end{center}
\end{minipage}
\caption{The current AERA-124 configuration on the left and an example of a horizontal event on the right.}
\label{fig:minipage_sol}
\end{figure}

AERA was designed as an engineering array and therefore there are two antenna types deployed in the AERA site. AERA-24 consists of 24 LPDAs, log periodic dipole antennas. These were deployed in November 2011. In May 2013 100 butterfly antenna stations were deployed and so AERA-124 was completed. A map of the current AERA configuration can be found in figure~\ref{fig:minipage_sol} on the left. The newly installed five butterfly stations with whisk-type antennas and the four tripole stations will be discussed in the next section.
%\begin{minipage}[b] take b or t

To determine the potential of radio detection of horizontal air showers a method was developed looking for the single station efficiency of AERA. Requiring an external trigger of the surface detector, the AERA events are selected respective to the distance of the closest AERA station to the shower axis and the timing of the strongest pulse in the AERA trace. The geometry of the shower is determined by the surface detector reconstruction in Offline~\cite{Offline}.

The quality cuts of the surface detector reconstruction for horizontal air showers are applied (ICRC 2013):
\begin{itemize}
\item zenith angle 62\(^\circ\)-80\(^\circ\)
\item minimum of four triggered surface detector stations, the surface detector stations have a distance of 1.5 km
%\item independent reconstruction of fluorescence detector energy
\item apply physics trigger cut, which rejects signals generated by background muons
\item apply quality trigger cut, which rejects events close to the border of the array
\item exclude bad periods
\end{itemize}

\begin{figure}[!htbp]
\includegraphics[height=.33\textheight]{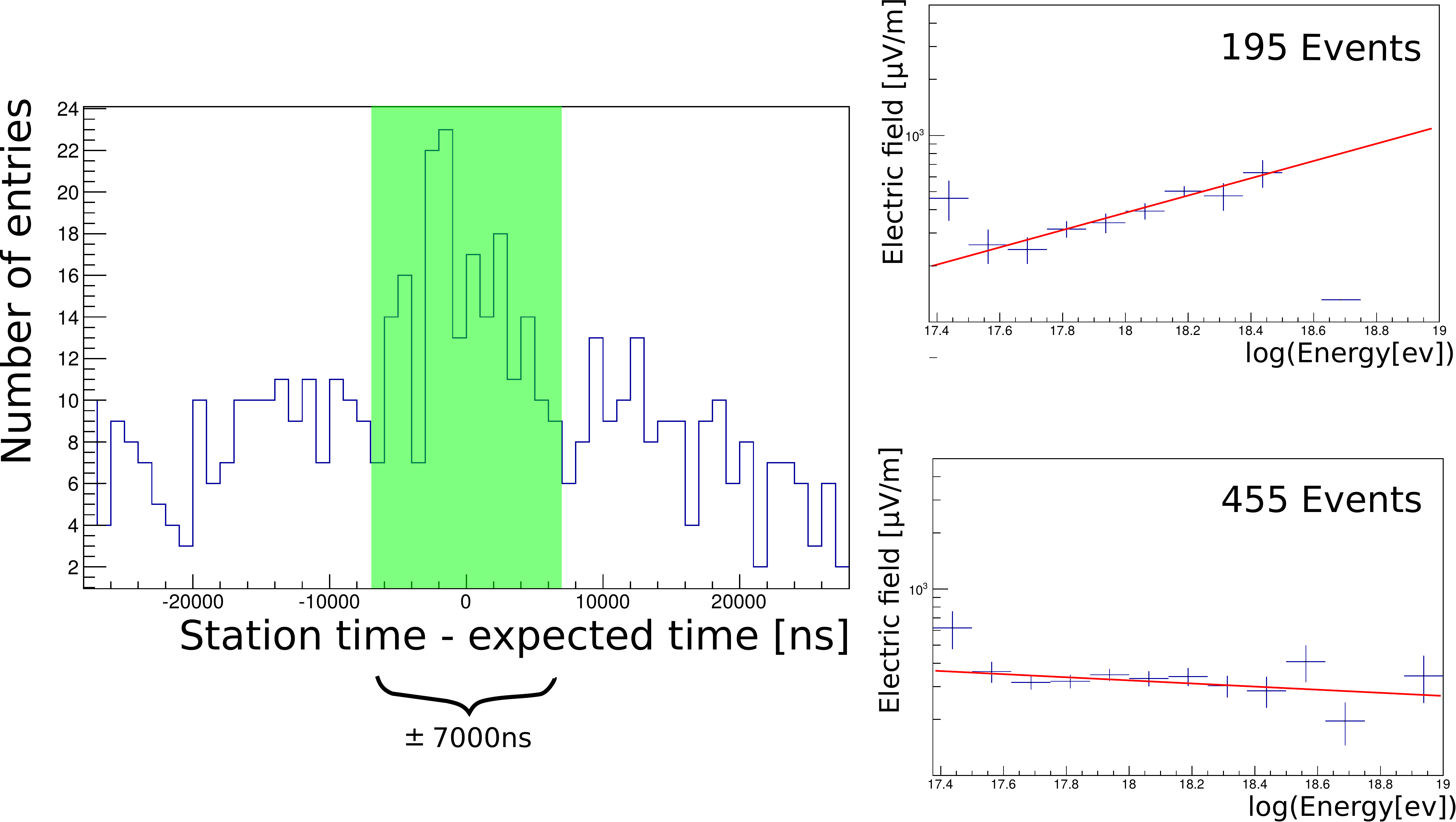}
\caption{Horizontal event selection. The distribution of station time of the pulse minus the calculated arrival time is shown on the left. The amplitude of events in the $\pm$ 7 \textmu s time interval are shown in the upper right corner. In the lower right corner events outside the $\pm$ 7 \textmu s time interval are shown. The fit was performed for the average of the amplitude of the three closest stations.}
\label{fig:timestamp}
\end{figure}

\begin{figure}[!htbp]
\includegraphics[height=.2\textheight]{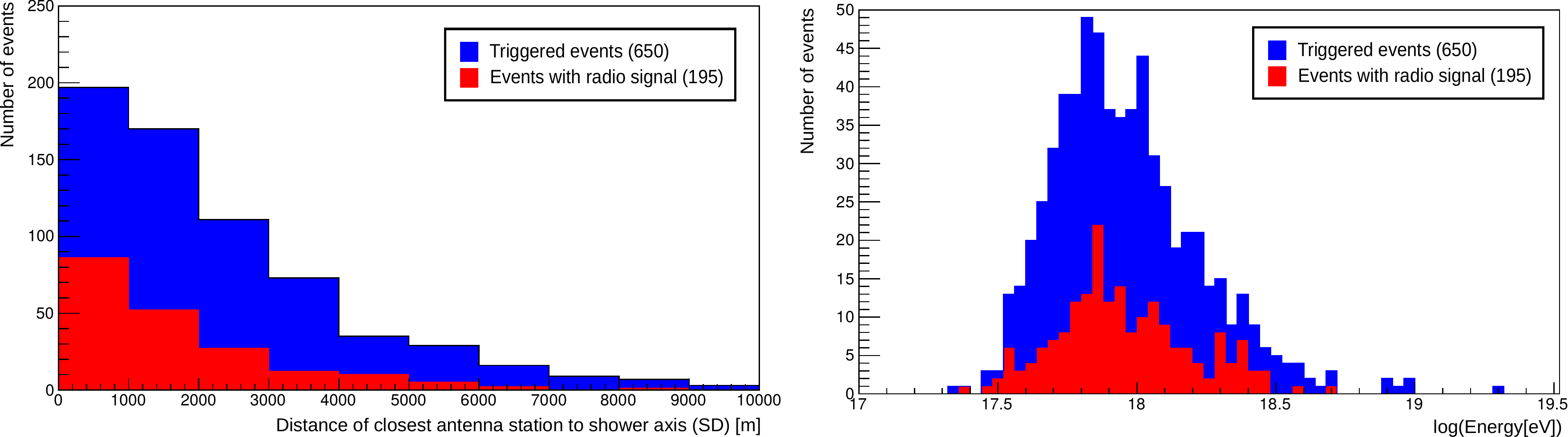}
\caption{The distance distribution (left) and the energy distribution (right) of selected events in the zenith angle range of 62\(^\circ\) - 80\(^\circ\) are shown.}
\label{fig:dist_energy}
\end{figure}

To distinguish between cosmic ray induced radio showers and a noise pulse the ratio of cosmic ray energy and the electric field strength measured by the AERA station is considered. The selection time interval for horizontal events is shown in figure~\ref{fig:timestamp}. The position of the magnitude of the electric field strength of the highest pulse of the closest AERA station is compared to the expected time in the trace. An interval of $\pm$ 7 \textmu s is taken as an event selection criterion. Currently the time interval is set that large due to the missing AERA time calibration and the large core uncertainty of horizontal air showers of low energy at the threshold. As the time calibration is established a sharper time window can be chosen. 
Events inside the time interval show a clear correlation between energy of the shower and electric field strength. This can be seen in figure~\ref{fig:timestamp} on the upper right corner. For events outside this time interval there is no correlation observed, which can be seen in figure~\ref{fig:timestamp} on the lower right corner. The fit was performed for the average of the amplitude of the three closest stations.
One example of a horizontal air shower is shown in figure~\ref{fig:minipage_sol} on the right. The event has an energy of 9.0$\times$10$^{17}$ eV, zenith angle of 64.5$^{\circ}$, azimuth angle of 323.6$^{\circ}$ and geomagnetic angle alpha of 101.4$^{\circ}$.
Figure~\ref{fig:dist_energy} shows the selected events when applying the selection criteria described above. On the left the distance distribution and on the right the energy distribution of the events are shown. Inclined showers can be seen up to a far distance to the reconstructed shower axis. \iffalse However, having confirmed experimentally the large footprint, investigations are still required to improve the purity and efficiency of the event selection since some high energy inclined events are not reconstructed by the present, not optimized, event selection.\fi
Although confirming experimentally the large footprint, there is still need of further investigations in order to improve the purity and the efficiency of the event selection since some high energy inclined events are not reconstructed by the present, not optimized, event selection. The new 25 AERA stations, which are going to be installed in 2015, will improve the statistic of horizontal air showers for AERA.

\section{Investigation of the vertical component}
\label{sec:vertical}

Simulation results have shown that for the reconstruction of inclined air showers the vertical electric field component becomes important. To investigate and improve the sensitivity of AERA to inclined showers, prototype stations including tripole antenna stations and whisk-type antennas as enhancement of the butterfly antennas have been deployed on the AERA site in November 2013. To study the emission in lower frequencies around 1 MHz one low-frequency tripole station has been deployed as well.
The current LPDAs and butterfly antenna stations measure two electric field components (North-South and East-West). The existing antenna technique is extended by a measurement of the vertical component of the electric field with nine newly installed prototype stations. Key aspects are the sensitivity to all incoming directions, finding the optimal frequency range and the coverage of a large area at low cost. Concerning inclined showers this requires an improved suppression of noise. This as well as the contribution of the radio emission to the vertical polarisation will be tested with the installed prototype stations.

%%%%%%%%%%%%%%%%%%%%%%%%%%%%%%%%%%%%%%%%%%%%%%%%%%%%%%%%%%%%%%%%%%%%%%%%%%%%%%%%%%%%%%%%%%%%%%%%%%%%%%%%%%%%%%%%%%

\section{Conclusion and outlook}

The radio detection of horizontal air showers has a lot of potential. The analysis of horizontal air showers will increase the number of events detected by AERA and will make a wider range of the sky visible. Simulations results show that the efficiency of detecting events rises with the zenith angle, this has to be shown for data too. AERA data of a period from 01/01/2012 till 19/03/2014 were analysed and first event candidates were selected. The selected events show a clear correlation between the primary shower energy and the electric field strength. Five whisk-type antennas and four tripole stations were deployed to determine the significance of the vertical component of the electric field vector. \\

%%%%%%%%%%%%%%%%%%%%%%%%%%%%%%%%%%%%%%%%%%%%%%%%%%%%%%%%%%%%%%%%%%%%%%%%%%%%%%%%%%%%%%%%%%%%%%%%%%%%%%%%%%%%%%%%%%

%%%%%%%%%%%%%%%%%%%%%%%%%%%%%%%%%%%%%%%%%%%%%%%%
%% BACKMATTER
%%%%%%%%%%%%%%%%%%%%%%%%%%%%%%%%%%%%%%%%%%%%%%%%

\noindent \small{{\bf Acknowledgement:} The corresponding author acknowledges the support through the second ASPERA~\citep{Aspera} call for R\&D, AugerNext~\citep{ AugerNext} and through the Helmholtz Alliance for Astroparticle Physics, HAP~\citep{HAP}.}

\bibliographystyle{aipproc}   % if natbib is available

\end{document}